\documentstyle[12pt,a4]{article}
\begin{document}
\makeatletter
\def\siml{\mathrel{\mathpalette\gl@align<}}
\def\simg{\mathrel{\mathpalette\gl@align>}}
\def\gl@align#1#2{\lower.6ex\vbox{\baselineskip\z@skip\lineskip\z@
 \ialign{$\m@th#1\hfill##\hfil$\crcr#2\crcr{\sim}\crcr}}}
\makeatother
\hbadness=10000
\hbadness=10000
\begin{titlepage}
\nopagebreak
\def\thefootnote{\fnsymbol{footnote}}
\begin{flushright}
{\normalsize
 DPSU-96-12\\
 INS-Rep-1161\\
September, 1996   }\\
\end{flushright}
\vspace{1cm}
\begin{center}
\renewcommand{\thefootnote}{\fnsymbol{footnote}}
{\large \bf Specific Scalar Mass Relations\\
in $SU(3) \times SU(2) \times U(1)$ Orbifold Model}

\vspace{1cm}

{\bf Yoshiharu Kawamura $^a$ 
\footnote[1]{e-mail: ykawamu@gipac.shinshu-u.ac.jp}}, 
{\bf Tatsuo Kobayashi $^b$
\footnote[2]{e-mail: kobayast@ins.u-tokyo.ac.jp}}\\
and\\
{\bf Tohru Komatsu $^a$}

\vspace{1cm}
$^a$ Department of Physics, Shinshu University \\

   Matsumoto, 390 Japan \\
and\\
$^b$ Institute for Nuclear Study, University of Tokyo \\
   Midori-cho, Tanashi, Tokyo, 188 Japan \\

\end{center}
\vspace{1cm}

\nopagebreak

\begin{abstract}
We study flat directions and soft scalar masses
using a $Z_3$ orbifold model with $SU(3) \times SU(2) \times U(1)$
gauge group and extra gauge symmetries including
an anomalous $U(1)$ symmetry.
Soft scalar masses contain $D$-term contributions and 
particle mixing effects after symmetry breaking and they are
parametrized by a few number of parameters.
Some specific relations among scalar masses are obtained.
\end{abstract}

\vfill
\end{titlepage}
\pagestyle{plain}
\newpage
\section{Introduction}
\renewcommand{\thefootnote}{\fnsymbol{footnote}}

Superstring theories (SSTs) are powerful candidates for the unification
theory of all forces including gravity.
There are various approaches to explore 4-dimensional (4-D) string models, 
for example, the compactification on Calabi-Yau manifolds \cite{CY}, 
the construction of orbifold models \cite{Orb,4DST} and so on.
Effective supergravity theories (SUGRAs) have been derived
by taking field theory limit \cite{ST-SG,OrbSG,OrbSG2}.

Effective low-energy theories have been derived under the assumption that 
supersymmetry (SUSY) is broken by $F$-term condensations of the 
dilaton field $S$ and/or moduli fields $T$ \cite{IL,ST-soft,ST-soft2}
from effective SUGRAs.
Some phenomenologically interesting features are predicted from 
the structure of soft SUSY breaking terms which
are parameterized by a few number of parameters, for example,
only two parameters such as a goldstino angle $\theta$ and the gravitino 
mass $m_{3/2}$ in the case with the overall moduli and the vanishing 
vacuum energy \cite{BIM}.
The cases with multimoduli fields are also discussed 
in Refs.\cite{multiT}.
Recently study on soft scalar masses has been extended
in the presence of an anomalous $U(1)$ symmetry \cite{N,U1A,U1X}.

Now we have thousands of effective low-energy theories
corresponding to 4-D string models following the above approach.
It is much important to select a realistic string model by some
experiments.
Soft SUSY breaking parameters can be powerful probes.
For example, string models with the SUSY breaking due to dilaton
$F$-term lead to the highly restricted pattern
such as \cite{ST-soft,BIM,Sdomi}
\begin{eqnarray}
|A| = |M_{1/2}| = \sqrt{3} |m_{3/2}| 
\label{dilaton}
\end{eqnarray}
where $A$ is a universal $A$-parameter, and gauginos and scalars get masses 
with common values $M_{1/2}$ and $m_{3/2}$, respectively.

In a recent paper \cite{K&K2}, the formula of soft SUSY breaking 
scalar masses has been derived from 4-D string models with 
flat directions \cite{flat}
within a more generic framework.
The effects of extra gauge symmetry breakings, 
that is, $D$-term and $F$-term contributions, particle mixing effects
and heavy-light mass mixing effects are considered.
The above prediction (\ref{dilaton}) does not hold in string models 
with an anomalous $U(1)$ symmetry \cite{U1A,K&K2}.

The purpose of this paper is to apply the formula to a 
semi-realistic string model including gauge groups and particle contents of
the minimal supersymmetric standard model (MSSM)
and to explore some excellent features.
Using a $Z_3$ orbifold model, 
we study flat directions and calculate soft scalar masses incorporating 
$D$-term contributions and particle mixing effects.
Specific relations among scalar masses are obtained.

This paper is organized as follows.
In the next section, we review the formula of soft SUSY breaking 
scalar masses derived from 4-D string models. 
In section 3, we study flat directions and
soft scalar masses using a $Z_3$ orbifold model.
Section 4 is devoted to conclusions and discussions.

\section{Formula of soft scalar masses}

We assume the existence of a realistic effective 
SUGRA, that is, our starting theory has the following
excellent feature.

The gauge group is 
$G=G_{SM} \times U(1)^n \times U(1)_A \times H'$ where 
$G_{SM}$ is the standard model gauge group 
$G_{SM} = SU(3)_C \times SU(2)_L \times U(1)_Y$, 
$U(1)^n$ are anomaly-free, 
$U(1)_A$ is anomaluos 
and $H'$ is a direct product of some 
non-abelian symmetries.
The anomalies related to $U(1)_A$ are canceled by 
the Green-Schwarz mechanism \cite{GS}.

Chiral multiplets are classified into two categories.
One is a set of chiral multiplets whose scalar components $\phi^i$ have
large VEVs of $O(M)$.
Here $M$ is the gravitational scale defined as 
$M \equiv M_{Pl}/\sqrt{8\pi}$ and $M_{Pl}$ is the Planck scale.
The dilaton field $S$ and the moduli fields $T_{i}$ belong to $\{\Phi^i\}$.
We treat only the overall moduli field $T$.
It is assumed that SUSY is broken by $F$-term condensations
of $\Phi^i$ such that 
$\langle F^i \rangle = O(m_{3/2}M)$.
The other is a set of matter multiplets denoted as $\Phi^{\kappa}$
which contains the MSSM matter multiplets and Higgs multiplets.
We denote the above two types of multiplets as $\Phi^I$ together.

We suppose the following situations related to extra gauge symmetry
breaking.
\begin{enumerate}
\item The $U(1)_A$ symmetry is broken by VEVs of $S$ and
some chiral matter multiplets.

\item Some parts of $U(1)^n$ and $H'$ are broken at much higher energy 
scales of $O(M_I)$ than the weak scale by VEVs of some chiral matter 
multiplets $\Phi^{\kappa}$.
Those VEVs are smaller than those of $S$ and $T$, i.e.
\begin{eqnarray}
\langle \phi^{\kappa} \rangle \ll \langle S \rangle, 
\langle T \rangle = O(M) .
\label{<phi>}
\end{eqnarray}
This condition is justified from the fact that a $D$-term condensation
of $U(1)_A$ vanishes up to $O(m_{3/2}^2)$.

\item Other extra gauge symmetries are broken spontaneously or 
radiatively by SUSY breaking effects at lower scales.
\end{enumerate}

In general, the fields diagonalizing SUSY mass terms, $\hat{\phi}^\kappa$, 
are given as linear combinations of original string states $\phi^\lambda$
such as 
\begin{eqnarray}
\hat{\phi}^\kappa = R^\kappa_\lambda \phi^\lambda .
        \label{R}
\end{eqnarray}
Coefficients $R^\kappa_\lambda$'s depend on
VEVs of moduli fields.

Assuming the vanishing cosmonogical constant, 
we have the following mass formula for light scalar fields $\hat{\phi}^k$ 
at the energy scale $M_I$ \cite{K&K2},
\begin{eqnarray}
(m^2)_k^l|_{M_I} &=& 
m_{3/2}^2  \delta_k^l
+ m_{3/2}^2 \cos^2\theta \hat{N}_{k}^l
\nonumber\\
&+& (V_{\rm Soft~Mass}^{\rm D})_k^l
+ (V_{\rm Soft~Mass}^{\rm Extra~F})_k^l
+ (V_{\rm Soft~Mass}^{\rm Mix})_k^l 
+ (V_{\rm Soft~Mass}^{\rm Ren})_k^l 
\label{SoftMassFormula}
\end{eqnarray}
where $(V_{\rm Soft~Mass}^{\rm D})_k^l$, 
$(V_{\rm Soft~Mass}^{\rm Extra~F})_k^l$,
$(V_{\rm Soft~Mass}^{\rm Mix})_k^l$ and 
$(V_{\rm Soft~Mass}^{\rm Ren})_k^l$ are 
$D$-term contributions, extra $F$-term
contributions, the contributions due to heavy-light mass mixing and
contributions related to renormalization effects from $M$ to $M_I$,
respectively.
Here $\theta$ is the so-called goldstino angle paramtrizing the ratio 
of $F^S$ and $F^T$ and $\hat{N}_{k}^l$ is obtained as 
$\hat{N}_k^l = (R^{-1})_k^\nu n_\nu R_\nu^l$ by modular weights $n_\nu$ 
in the $\phi^\kappa$-basis.

$D$-term contributions play an important role for later discussions 
and are given as \cite{D-term,KMY2,Kawa1}
\begin{eqnarray}
(V_{\rm Soft~Mass}^{\rm D})_k^l &=& \sum_\alpha g^2_\alpha 
\langle D^\alpha \rangle (\hat{Q}^\alpha)_k^l 
\label{SoftMassD}
\end{eqnarray}
where $g_\alpha$'s are gauge coupling constants and
\begin{eqnarray}
(\hat{Q}^\alpha)_k^l &\equiv& \langle \hat{K}_k^\mu \rangle 
(\hat{q}^\alpha)_\mu^l ,~~~~
(\hat{q}^\alpha)_\kappa^\lambda \equiv (R^{-1})_\kappa^{\nu}
          q^\alpha_{\nu} R_{\nu}^\lambda .
\end{eqnarray}
Here $\hat{K}_I^J$ is a K\"ahler metric and $q_\nu^\alpha$'s are
diagonal charges. 
The $D$-term condensations are written as
\begin{eqnarray}
g^2_{\hat{\alpha}} \langle D^{\hat{\alpha}} \rangle &=& 2 
g_{\hat{\alpha}} m_{3/2}^2 
\{ (M_V^{-2})^{\hat{\alpha} A} g_A (1 - 6C^2 \sin^2\theta )
\langle \sum_\kappa {q_\kappa^A} (T+T^*)^{n_\kappa} |\phi^\kappa|^2 \rangle
\nonumber \\
&~& ~~~~ -  \sum_{\hat{\beta}} (M_V^{-2})^{\hat{\alpha}\hat{\beta}}
g_{\hat{\beta}} C^2 \cos^2\theta 
\langle \sum_\kappa q_\kappa^{\hat{\beta}} n_\kappa (T+T^*)^{n_\kappa}
 |\phi^\kappa|^2 \rangle \} 
\label{D-cond2}
\end{eqnarray}
where $(M_V^{-2})^{\alpha\beta}$ is the inverse matrix of
gauge boson mass matrix $(M_V^2)^{\alpha\beta}$ given as
\begin{eqnarray}
(M_V^2)^{\alpha\beta} = 
  2 g_{\alpha} g_{\beta} \langle (T^\beta(\phi^\dagger))_I
K^I_J (T^\alpha(\phi))^J \rangle .
        \label{MV2}
\end{eqnarray}
Here the gauge transformation of $\phi^I$ is 
given as $\delta \phi^I = i g_{\alpha} (T^\alpha (\phi))^I$
up to space-time dependent infinitesimal parameters.
Here indices $\hat{\alpha}$ and $\hat{\beta}$ run over broken generators.
Further the gaugino mass is obtained as \cite{BIM}
\begin{eqnarray}
M^2_{1/2}=3m_{3/2}^2\sin^2 \theta.
\label{gmass}
\end{eqnarray}

\section{Examples}

\subsection{Flat direction}

We study the $Z_3$ orbifold model with 
a shift vector $V$ and Wilson lines $a_1$ and $a_3$ such 
as \cite{stringmodel,flat22}
$$V={1 \over 3}(1,1,1,1,2,0,0,0)(2,0,0,0,0,0,0,0),$$
$$a_1={1 \over 3}(0,0,0,0,0,0,0,2)(0,1,1,0,0,0,0,0),$$
$$a_3={1 \over 3}(1,1,1,2,1,0,1,1)(1,1,0,0,0,0,0,0).$$
This model has a gauge group as 
$ G = SU(3)_C \times SU(2)_L \times U(1)^8 \times SO(10)'$.
The $U(1)$ charge generators are defined in Table 1 
and one of them is anomalous.

This model has matter multiplets as 
$${\rm U-sec.} :\quad 3[(3,2)_0+(\bar{3},1)_0 +(1,2)_0]
+3(16)'_9,$$
\begin{eqnarray*}
{\rm T-sec.} &: \quad &3[4(3,1)_4+5(\bar{3},1)_4] 
+3[11(1,2)_4+(1,2)_{-8}]\\
(N_{OSC}=0)&~& +114(1,1)_4+30(1,1)_{-8},
\end{eqnarray*}
$${\rm T-sec.} (N_{OSC}=-1/3):\quad 27(1,1)_4$$
where the number of suffix denotes the anomalous $U(1)$ charge 
and $N_{OSC}$ is the oscillator number.
Thus there are many $G_{SM}$-singlets in this model.
In particular, the following 
$G_{SM}$-singlets play an important role for study of flat directions 
leading to realistic vacua
\begin{eqnarray}
& S_1: \ & Q_a= (-6,0,0,2,0,4,0,-8),\nonumber \\
& S_2: \ & Q_a= (0,-4,0,-2,-2,0,4,-8),\nonumber \\
& S_3: \ & Q_a= (0,-4,0,-2,2,-4,-4,-8),\nonumber \\
& S_6: \ & Q_a= (6,4,0,0,-2,0,-2,-8),\nonumber \\
& S_8: \ & Q_a= (6,4,0,0,2,2,2,-8),\nonumber \\
& S_{10}: \ & Q_a= (-6,0,0,2,-4,-4,-4,4),\nonumber \\
& S_{11}: \ & Q_a= (-6,0,0,2,4,0,4,4),\nonumber \\
& Y_1: \ & Q_a= (-6,0,0,2,0,-2,0,4),\nonumber \\
& Y_3: \ & Q_a= (0,-4,0,-2,2,2,2,4)\nonumber \\
& A_5: \ & Q_a= (-3,-2,3,3,1,0,-2,-8)\nonumber \\
& \overline A_5: \ & Q_a= (-3,-2,-3,-3,1,0,-2,-8)\nonumber 
\end{eqnarray}
where  $Q_a$ $(a=1,2,...,7,A)$ are $U(1)$ charges
and we follow the notation of the fields in Ref.~\cite{stringmodel}. 
The fields $S_i$, $A_i$ and $\overline A_i$ correspond 
to the non-oscillated twisted sector 
with $n_\kappa =-2$ and $Y_j$ corresponds to the twisted sector with a 
nonvanishing oscillator number.
Thus the fields $Y_j$ have the modular weight $n_Y=-3$.
There exist several types of flat directions 
in the SUSY limit \cite{stringmodel,flat22}.

Let us take one example where the flat direction is given as 
\cite{stringmodel}
\begin{eqnarray}
&~&\langle (T+T^*)^{-3}|Y_3|^2 \rangle =  v, \nonumber \\
&~&\langle (T+T^*)^{-3}|Y_1|^2 \rangle =  
\langle (T+T^*)^{-2}|S_6|^2 \rangle =  u+v,\nonumber \\
&~&\langle (T+T^*)^{-2}|S_1|^2 \rangle =  
\langle (T+T^*)^{-2}|S_2|^2 \rangle =  
\langle (T+T^*)^{-2}|S_3|^2 \rangle =  \nonumber \\
&~&\langle (T+T^*)^{-2}|S_8|^2 \rangle =  u
\end{eqnarray}
where $u$ and $v$ are positive constants which are determined
by the $D$-flatness condition and the minimum of scalar potential 
after SUSY breaking.
Along this flat direction, the $U(1)$ symmetries break 
as $U(1)^8 \rightarrow U(1)^2$.
One of unbroken $U(1)^2$ corresponds to $Q_3$.
The other is a linear combination of $Q_1$, $Q_2$ and $Q_4$ as 
$Q_1/3-Q_2/2+Q_4$ which is regarded as the hypercharge.

We define broken $U(1)$ charges as
\begin{eqnarray}
&~& Q'^1 \equiv {1 \over \sqrt{5}}(Q_1 + Q_2) ,~~~
Q'^2 \equiv {1 \over \sqrt{55}}(2Q_1 - 3Q_2 - 5Q_4) ,
\nonumber\\
&~& Q'^5 \equiv Q_5 ,~~~Q'^6 \equiv {1 \over \sqrt{2}}Q_6 ,~~~
Q'^7 \equiv {1 \over \sqrt{2}}Q_7 ,~~~Q'^A \equiv {1 \over \sqrt{3}}Q_A .
\label{Qprime}
\end{eqnarray}
Note that the gauge boson mass matrix is not diagonalized 
in this definition.
The modular weights and broken $U(1)$ charges of 
the fields with VEVs are given in Table 2.

The $D$-flatness condition for $U(1)_A$ requires
\begin{eqnarray}
\langle {\delta_{GS} \over S+S^*} \rangle -36u=0
\end{eqnarray}
where $\delta_{GS}^{A}$ is a coefficient of the Green-Schwarz
mechanism \cite{GS} to cancel the $U(1)_A$ anomaly and 
is given as 
\begin{eqnarray}
  \delta_{GS}^{A} &=& {1 \over 96\pi^2}Tr Q_A .
\label{delta_GS}
\end{eqnarray}
In addition, we have 
\begin{eqnarray}
f(n_\kappa^2)=29u+22v
\end{eqnarray}
where $f(a_\kappa)$ is defined as
\begin{eqnarray}
f(a_\kappa)=\langle \sum_\kappa a_\kappa (T+T^*)^{n_\kappa} 
|\phi^{\kappa}|^2 \rangle .
\end{eqnarray}
The scalar potential includes $f(n_\kappa^2)$ \cite{K&K2}.
The minimum of scalar potential 
is obtained at the following point:
\begin{eqnarray}
u ={1 \over 36}\langle {\delta_{GS}^A \over S+S^*} \rangle, \quad 
v \sim O(m_{3/2}^2) .
\label{vi}
\end{eqnarray}
Using $Tr Q_A=1296$ and $\langle ReS \rangle \sim 2$, 
we estimate $u \sim M^2/105$. From Eqs.(\ref{vi}), 
the breaking scale of $U(1)'_i$'s is estimated as $M_I=O(u^{1/2})$.

Along this flat direction, several fields gain mass terms.
For example, we consider mass terms among $(3,1)$ and 
$(\overline 3,1)$ fields in the twisted sector.
Here we follow the notation of fields in Ref. \cite{stringmodel}.
The $D_1$ field appears as a massless $(3,1)$ field in the 
twisted sector with the Wilson line $(m_1,m_3)=(0,0)$, where 
$(m_1,m_3)$ denotes the Wilson line $a=m_1a_1+m_3a_3$.
Further $(\overline 3,1)$ fields include the $d_1$ and $d_2$ fields, 
which have $(m_1,m_3)=(0,1)$ and $(-1,1)$, respectively.
Selection rules due to space group invariance allow couplings satisfying 
the following condition:
\begin{eqnarray}
\sum m_1=3\ell, \quad \sum m_3=3\ell'
\label{SGinv}
\end{eqnarray}
where $\ell$ and $\ell'$ are integers.
Further its coupling strength is obtained for each $i$-th plane 
as $h_i \sim e^{-a\langle T_i \rangle}$ 
where $a=0$ 
in the case that $m_i$ takes a same number for all species
and $a>0$ for other cases \cite{couple}.
The singlet fields $S_2$ and $S_3$ have $(m_1,m_3)=(0,-1)$ and 
$(1,-1)$, respectively.
The space group invariance (\ref{SGinv}) as well as gauge invariance 
allows the following couplings:
\begin{eqnarray}
D_1d_1S_2, \quad D_1d_2S_3.
\end{eqnarray}
These couplings include suppression factors as $h_2$ and 
$h_1 h_2$, respectively.
Their mass terms along the flat direction are written as 
\begin{eqnarray}
\langle S_2 \rangle h_2(d_1
+h_1{\langle S_3 \rangle \over \langle S_2 \rangle}d_2)D_1.
\end{eqnarray}
The light field is obtained as the linear combination 
$h_1d_1-d_2$, up to normalization.
Thus the matrix $R^\lambda_\kappa$ involves the 
moduli-dependent function $h_i$.\footnote{
This moduli-dependence of the diagonalizing matrix has not been 
discussed in Ref.\cite{stringmodel}.}
Furthermore, the matrix $R^\lambda_\kappa$ is, in general, 
dependent of ratios of VEVs.
The other $SU(3)$ triplets fields in the twisted sector become massive.
Similarly we obtain  the Higgs field $H$ by string states $G_1$ and 
$G_2$ in the twisted sector as $h_1G_2-G_3$.
Here $H$ and ${\overline{H}}$ are the Higgs
doublets with hypercharge $-1/2$ and $1/2$, respectively.
The other MSSM matter fields coincide with the string states.
If we take into account nonrenormalizable couplings, 
up-type quarks are obtained as linear combinations 
of string states \cite{stringmodel}.
However, mass terms induced by nonrenormalizable couplings include 
suppression factors of $O((u/M)^{1/2}) \sim 1/10$.
Here we neglect such effects.

This orbifold model has another flat direction as \cite{flat22}
\begin{eqnarray}
&\langle (T+T^*)^{-2}|S_6|^2 \rangle =  
2\langle (T+T^*)^{-2}|S_{11}|^2 \rangle = 4 \lambda a,\nonumber \\
&\langle (T+T^*)^{-2}|\overline{A_5}|^2 \rangle =  
\langle (T+T^*)^{-2}|A_5|^2 \rangle = (2+2\lambda )a, \nonumber \\
&\langle (T+T^*)^{-2}|S_8|^2 \rangle =  
2\langle (T+T^*)^{-2}|S_{10}|^2 \rangle = (8-4 \lambda )a, \nonumber \\
&\langle (T+T^*)^{-2}|S_1|^2 \rangle =  
\langle (T+T^*)^{-2}|S_3|^2 \rangle = (2-2\lambda )a, \nonumber \\
&\langle (T+T^*)^{-2}|S_2|^2 \rangle = 4a
\label{model2}
\end{eqnarray}
where $a$ and $\lambda$ satisfy 
\begin{eqnarray}
\langle {\delta_{GS} \over S+S^*} \rangle -18a=0, \quad 
0 \leq \lambda \leq 1.
\label{model21}
\end{eqnarray}
Along this flat direction we have 
$f(n_\kappa^2) = 48a$.
It is notable that $f(n_\kappa^2)$ is independent of $\lambda$.
Thus the direction corresponding to the parameter $\lambda$ 
is still a flat direction at this level, although 
this vacuum has a larger $f(n_\kappa^2)$ than the previous one.

\subsection{Scalar mass relations}

Let us calculate soft scalar masses using the formula
(\ref{SoftMassFormula}) and derive specific relations among them.
The basic idea and strategy are the same as those 
in Refs.\cite{KMY1,KMY2,KT}.
The SUSY spectrum at the weak scale, which is expected to 
be measured in the near future, is translated into the soft 
SUSY breaking parameters.
The values of these parameters at higher energy scales 
are obtained by using the renormalization 
group equations (RGEs) \cite{RGE}.
In many cases, there exist 
some relations among these parameters. 
They reflect the structure of high-energy physics.
Hence we can specify the high-energy physics by examining
these relations.

We have the same number of observable soft masses as that of species
of scalar fields and gauginos.
There are several unknown parameters in the RHS of 
Eq.(\ref{SoftMassFormula}) such as $m_{3/2}^2$ and $\cos^2\theta$.
If the number of independent equations is more than that of unknown
parameters, non-trivial relations exist among soft masses.
They can be obtained by eliminating unknown parameters.

In our model, the breaking scale $M_I$ is 
estimated as $O(10^{-1} M)$ and so renormalization effects 
from $M$ to $M_I$ are neglected.
We assume that Yukawa couplings among heavy and light fields are 
small enough and the $R$-parity is conserved.
In such a case, we can neglect the effect of extra $F$-term 
contributions.
Since there are no sizable mixing terms among heavy and 
light fields in the K\"ahler potential in $Z_3$ orbifold models
as shown in appendix A,
there appear no heavy-light mixing terms
of $O(m_{3/2} M_I)$ if Yukawa couplings among heavy, light and moduli
fields are suppressed sufficiently, i.e., $O(m_{3/2}/M)$.
We assume that string state mixing occurs among the same generation
after the breakdown of extra gauge symmetries, that is,
there is no flavor mixing.

Under the above assumptions,
our soft scalar mass formula is written in a simple form such as
\begin{eqnarray}
m^2_k|_{M_I} &=& m_{3/2}^2 + m_{3/2}^2  \hat{N}_k \cos^2\theta 
+ \sum_{\hat{\alpha}} g_{\hat{\alpha}}^2
\langle D^{\hat{\alpha}} \rangle (\hat{Q}^{\hat{\alpha}})_k
\label{m2_k}
\end{eqnarray}
where 
$\hat{N}_k = (R^{-1})_k^\nu n_\nu R_\nu^k$ and 
$(\hat{Q}^\alpha)_k = (R^{-1})_k^\nu q_\nu^{\alpha} R_\nu^k$ 
without summations for $k$.
In Table 3, the modular weights and broken $U(1)$ charges
for light scalar fields are given.

The gauge boson mass matrix is represented as
$(M_V^2)^{\alpha\beta} = 2 g'_{\alpha} g'_{\beta} f(q'^{\alpha}q'^{\beta})$.
Here $g'_{\alpha}$'s are gauge coupling constants defined in the basis
(\ref{Qprime}) and $q'^{\alpha}_k$'s represent $U(1)$ charges 
$(\hat{Q}'^{\alpha})_k$ for scalar fields $\hat{\phi}^k$.
The $D$-term condensations are written as
\begin{eqnarray}
g^2_{\hat{\alpha}} \langle D^{\hat{\alpha}} \rangle &=&  
m_{3/2}^2 f^{-1}(q'^{\hat{\alpha}}q'^{\hat{\beta}}) U^{\hat{\beta}}
\label{D-cond3}
\end{eqnarray}
where $f^{-1}(q'^{\hat{\alpha}}q'^{\hat{\beta}})$ is 
the inverse matrix of $f(q'^{\hat{\alpha}}q'^{\hat{\beta}})$
and $U^{\hat{\beta}}$ is given as\footnote{
Here we assume that the values of all gauge couplings equal at $M_I$.}
\begin{eqnarray}
U^{\hat{\beta}} = (1-6\sin^2\theta)f(q'^{\hat{\beta}}) 
\delta^{A}_{\hat{\beta}}
- \cos^2\theta  f(n_\kappa q'^{\hat{\beta}}) .
\label{U}
\end{eqnarray}

We need the values for $f(q'^{\hat{\alpha}})$, 
$f(n_\kappa q'^{\hat{\alpha}})$ and 
$f(q'^{\hat{\alpha}}q'^{\hat{\beta}})$
to calculate $D$-term contributions.
The values for $f(q'^{\hat{\alpha}})$ and $f(n_\kappa q'^{\hat{\alpha}})$
are calculated as 
\begin{eqnarray}
&~&f(q'^{\hat{\alpha}}) = 0 ~~(\hat{\alpha} = 1,2,5,6,7),~~
f(q'^A) = -12\sqrt{3}u,
\label{fq}\\
&~&f(n_{\kappa}q'^1) = {6u+10v \over \sqrt{5}},~~
f(n_{\kappa}q'^2) = {2 \over 5}\sqrt{55}u,~~
f(n_{\kappa}q'^5) = -2v,
\nonumber\\
&~&f(n_{\kappa}q'^6) = \sqrt{2}u,~~
f(n_{\kappa}q'^7) = -\sqrt{2}v,~~
f(n_{\kappa}q'^A) = {68u-8v \over \sqrt{3}}.
\label{fnq}
\end{eqnarray}
Using the above values,
$U^{\hat{\beta}}$ is calculated as
\begin{eqnarray}
(U^{\hat{\beta}})^T &=& (-{6 \over 5}\sqrt{5}\cos^2\theta,
-{2 \over 5}\sqrt{55}\cos^2\theta, 0, -{\sqrt{2}}\cos^2\theta,
\nonumber \\
&~&~~~~   0, {1 \over \sqrt{3}}(180 - 284\cos^2\theta)) u .
\label{U2}
\end{eqnarray}
The values for $f(q'^{\hat{\alpha}}q'^{\hat{\beta}})$ are calculated as 
\begin{eqnarray}
f(q'^{\hat{\alpha}}q'^{\hat{\beta}}) =
\left(
\begin{array}{cccccc}
{304 \over 5} & {8\sqrt{11} \over 5}& 0 & {12\sqrt{10} \over 5}
& 0 & -{24\sqrt{15} \over 5}\\
{8\sqrt{11} \over 5} & {176 \over 5} & 0 & -{6\sqrt{110} \over 5}
& 0 & -{8\sqrt{165} \over 5}\\
0 & 0 & 16 & -2\sqrt{2} & -4\sqrt{2} & 0 \\
{12\sqrt{10} \over 5} & -{6\sqrt{110} \over 5} & -2\sqrt{2} & 20
& 10 & -4\sqrt{6}\\
0 & 0 & -4\sqrt{2} & 10 & 20 & 0 \\
-{24\sqrt{15} \over 5} & -{8\sqrt{165} \over 5} & 0 
& -4\sqrt{6} & 0 & 112
\end{array}
\right)u .
\label{fqq}
\end{eqnarray}
Using Eqs.(\ref{m2_k}), (\ref{D-cond3}), (\ref{U2}) and (\ref{fqq}), 
we have obtained soft scalar masses at $M_I$ in the following form,
\begin{eqnarray}
m^2_k|_{M_I} &=& m_{3/2}^2 (a + b \cos^2\theta) .
\label{m2_kab}
\end{eqnarray}
In Table 4, we give the values of $a$ and $b$ for all species.
The values $a$ and $a+b$ corresponds to
the extreme cases $\cos^2 \theta = 0$ and $\cos^2 \theta = 1$
for mass ratios $m^2_k/m_{3/2}^2|_{M_I}$, respectively.
Note that the $h_1$ dependence disappears.

Many fields can acquire negative squared masses and 
they could trigger a \lq \lq larger'' symmetry breaking 
including the dangerous color and/or charge symmetry breaking.
The fifth column of Table 4 shows the range of $\cos^2 \theta$ leading 
to $m_k^2 \geq 0$ at the tree level for each sfermion.
Radiative corrections due to gaugino masses (\ref{gmass}) are 
important for squark masses.
The sixth column of Table 4 shows the range of $\cos^2 \theta$ leading 
to $m_k^2 \geq 0$ at $M_Z$ for each sfermion including one-loop radiative 
corrections.
Here we neglect RGE effects of Yukawa couplings.\footnote{
It is valid for the first and second families.}
All of the sfermions have $m_k^2 \geq 0$ at $M_Z$ in the range with 
$0.61 \leq \cos^2 \theta \leq 0.87$.
The $\mu$-term as well as the soft mass terms contributes to the 
Higgs mass terms.
Hence we omit the ranges leading to $m^2_{H(\overline H)} \geq 0$ for 
soft masses in the fifth and sixth columns of Table 4.
A suitable $\mu$-term could lead to a successful symmetry breaking.
Here we do not discuss the $\mu$-term explicitly since 
that is beyond this work.
In addition we have a strong non-universality of soft masses, i.e.  
$m_k^2=O(10m^2_{3/2})$ 
for some fields while $m_k^2=O(m^2_{3/2})$ for others.
Note that we have non-universal soft masses even in the case with 
$\cos \theta =0$.
That is a generic feature of models with anomalous $U(1)$ 
symmetry \cite{U1A,K&K2}.
As a feature of this model, soft masses are degenerate for squarks 
and sleptons with same quantum numbers under $G_{SM}$
because they have same quantum numbers under gauge group $G$ 
and same modular weights.
Hence the process of flavor changing neutral current (FCNC)
\cite{FCNC} is sufficiently suppressed.

Let us obtain relations among scalar masses and gaugino masses.
As we have eight kinds of observables ($m_{\tilde{q}}$, $m_{\tilde{u}}$,
$m_{\tilde{d}}$, $m_{\tilde{l}}$, $m_{\tilde{e}}$, $m_H$,
$m_{\overline{H}}$, $M_{1/2}$) and two unknown parameters
($m_{3/2}$, $\cos\theta$), we can obtain at least six
independent relations.
In fact, we have the following relations
\begin{eqnarray}
&~& 3(m_{\tilde{l}}^2 - m_{\tilde{e}}^2) 
   = m_{\tilde{u}}^2 - m_{\tilde{q}}^2 ,
\label{Massrelation1} \\
&~& m_{\tilde{q}}^2 + m_H 
   = m_{\tilde{u}}^2 + m_{\tilde{d}}^2 ,
\label{Massrelation2} \\
&~& m_{\tilde{q}}^2 + m_{\tilde{d}}^2 +
   4(m_{\tilde{u}}^2 + m_{\tilde{e}}^2) = 0 ,
\label{Massrelation3} \\
&~& 13(m_{\tilde{q}}^2 + m_{\tilde{d}}^2) +
   12(m_{\tilde{l}}^2 + m_H) = 0 ,
\label{Massrelation4} \\
&~& 2m_{\tilde{q}}^2 + 3m_{\tilde{d}}^2 
   + m_{\tilde{e}}^2  = m_{\overline{H}}^2 ,
\label{Massrelation5} \\
&~& m_{\tilde{q}}^2 + m_{\tilde{u}}^2 
   + m_{\overline{H}}^2  = M^2_{1/2}  
\label{Massrelation6}
\end{eqnarray}
where the tilde represents the scalar component.
Similarly we can obtained soft scarlar masses for other vacua, e.g. 
Eq.(\ref{model2}).

\section{Conclusions and Discussions}

We have studied flat directions and soft scalar masses
using a $Z_3$ orbifold model with $SU(3) \times SU(2) \times U(1)$
gauge group and extra gauge symmetries including
an anomalous $U(1)$ symmetry.
Soft scalar masses contain $D$-term contributions and particle mixing
effects after extra gauge symmetry breaking and they are
parametrized by a few number of parameters.

We have calculated soft scalar masses at $M_I$.
It is, in general, difficult to keep the degeneracy and 
positivity of squared masses at the tree level.
We have non-universal soft masses even for $\cos \theta =0$ 
as a generic feature of string models with anomalous $U(1)$ breaking.
This fact does not lead to serious problems for FCNC in our model.
Because soft masses are degenerate for squarks 
and sleptons with same quantum numbers under $G_{SM}$.
A strong non-universality of soft masses, i.e.  
$m_k^2=O(10m^2_{3/2})$ 
for some fields while $m_k^2=O(m^2_{3/2})$ for others
might provide interesting implications in the phenomenological
viewpoint.
In fact, much work is devoted to phenomenological implications
of the non-universality of soft masses \cite{nonuni}.
The positivity of $m^2$ can be recovered by radiative corrections.
It is an interesting subject to examine whether the radiative breaking 
scenario \cite{RGE} can be realized.

We have obtained some specific relations among scalar masses.
They can be powerful probes to specify a realistic model
based on 4-D string models.

\appendix

\section{Heavy-light Mixing in K\"ahler potential 
}
\label{app:A}

If the MSSM matter fields $\hat{\phi}_{(SM)}^k$ are given 
as linear combinations of string states with the same modular 
weight $n_k$, the K\"ahler potential of matter part is
given as
\begin{eqnarray}
K^{(M)} &=& \sum_{({\rm SM})} (T+T^*)^{n_k} |\hat{\phi}_{({\rm SM})}^k|^2 
+ \cdots .
\label{K(M)string'}
\end{eqnarray}
where the ellipses stand for terms related to fields other than the 
MSSM matter fields.
In this case, there are no heavy-light mixing terms
in $K^{(M)}$.
Whether the SM matter fields $\hat{\phi}_{({\rm SM})}^k$ are given as 
linear combinations of original fields with the same modular weight or not
is model-dependent.
We discuss this issue based on $Z_N$ orbifold models in this appendix.

The explicit model in section 3 shows the origin of particle mixing 
as follows.
Suppose that we have the following two couplings:
\begin{eqnarray}
\phi \phi_1 \prod_i \chi_i, \quad \phi \phi_2 \prod_j \chi'_j,
\end{eqnarray}
including the common field $\phi$.
On the top of that, we assume this model has flat directions as 
$\langle \prod_i \chi_i \rangle \neq 0$ and 
$\langle \prod_j \chi'_j \rangle \neq 0$.
Then mass eigenstates are obtained as linear combinations of $\phi_1$ and 
$\phi_2$.
If these fields, $\phi_1$ and $\phi_2$, have the same modular weight $n_k$, 
the light field among their linear combinations has its K\"ahler 
potentail as Eq.~(\ref{K(M)string'}).
Otherwise, its K\"ahler potential becomes complicated.

For example, we study $Z_3$ orbifold models.
These models have two types of renormalizable couplings as 
\begin{eqnarray}
\phi_{U1}\phi_{U2}\phi_{U3}, \quad 
\phi_{T1}\phi'_{T1}\phi''_{T1},
\end{eqnarray}
where $\phi_{Ui}$ denotes a field in one of the three untwisted sectors 
and $\phi_{T1}$ corresponds to that in the twisted sector.
These couplings have no common field.
Thus particle mixing with different modular weights does not 
appear at this level.
Nonrenormalizable couplings \cite{nonre} could lead 
to particle mixing, 
but these couplings include a suppression factor 
$(\langle \chi \rangle /M)^n$.
Thus such effects are negligible in most of cases.
In the same way, the particle mixing effect is negligible
in $Z_7$ orbifold models since it can appear only through
nonrenormalizable couplings.

$Z_{2n}$ orbifold models are different from $Z_3$ and $Z_7$
orbifold models.
Because $Z_{2n}$ orbifold models have several types of renormalizable 
couplings \cite{Yukawa}.
For example, $Z_4$ orbifold models have three types of 
renormalizable couplings as
\begin{eqnarray}
\phi_{T1}\phi'_{T1}\phi_{T2}, \quad \phi_{T2}\phi'_{T2}\phi_{U3},
\label{RCZ4}
\end{eqnarray}
in addition to $\phi_{U1}\phi_{U2}\phi_{U3}$.
Here $\phi_{T1}$ and $\phi_{T2}$ correspond to fields in the 
$\theta$-twisted and $\theta^2$-twisted sectors,
where $\theta$ denotes the $Z_4$-twist to construct $Z_4$ orbifold models.
We consider the couplings (\ref{RCZ4}) and assume the existence of 
flat directions as 
$\langle \phi_{U3}\rangle \neq 0$ and 
$\langle \phi_{T1}\rangle \neq 0$.
In this case mass eigenstates are linear combinations of $\phi'_{T1}$ 
and $\phi'_{T2}$, which have modular weights $n_k=-2$ and $-1$, 
respectively.
Therefore sizable particle mixing with different modular weights 
can appear in $Z_{2n}$ orbifold models.

\newpage

\section*{Table Captions}

\renewcommand{\labelenumi}{Table~\arabic{enumi}}
\begin{enumerate}
\item $U(1)$ charge generators in terms of $E_8 \times E_8'$ lattice
vectors.

\item The modular weights and broken $U(1)$ charges of
the scalar fields with VEVs.
We follow the notation of fields in Ref.\cite{stringmodel}.

\item The modular weights and broken $U(1)$ charges for light 
scalar fields.
Here $h_1$ denotes $h_1 \sim e^{-a\langle T_1 \rangle}$. 

\item The particle contents and the ratios of $m_k^2/m_{3/2}^2$.
We omit the ranges leading to $m^2_{H(\overline H)} \geq 0$ 
for soft masses.
\end{enumerate}

\newpage

\begin{center}
{\Large Table 1} 
\end{center}
\begin{eqnarray*}
Q_1 &=& 6(1,1,1,0,0,0,0,0)(0,0,0,0,0,0,0,0)'\\
Q_2 &=& 6(0,0,0,1,-1,0,0,0)(0,0,0,0,0,0,0,0)'\\
Q_3 &=& 6(0,0,0,0,0,1,0,0)(0,0,0,0,0,0,0,0)'\\
Q_4 &=& 6(0,0,0,0,0,0,1,0)(0,0,0,0,0,0,0,0)'\\
Q_5 &=& 6(0,0,0,0,0,0,0,1)(0,0,0,0,0,0,0,0)'\\
Q_6 &=& 6(0,0,0,0,0,0,0,0)(1,1,0,0,0,0,0,0)'\\
Q_7 &=& 6(0,0,0,0,0,0,0,0)(0,1,1,0,0,0,0,0)'\\
Q_A &=& 6(0,0,0,0,0,0,0,0)(1,-1,1,0,0,0,0,0)'
\end{eqnarray*}

~~\

\begin{center}
{\Large Table 2} 
\end{center}

\begin{tabular}{|c|c|c|c|c|c|c|c|}\hline
 String state & $n_k$ & $\sqrt{5}\hat{Q}'^1$ 
& $\sqrt{55}\hat{Q}'^2$ &
 $\hat{Q}'^5$ & $\sqrt{2}\hat{Q}'^6$ & $\sqrt{2}\hat{Q}'^7$ 
& $\sqrt{3}\hat{Q}'^A$ \\ \hline
 $S_1$ & $-2$ & $-6$ & $-22$ & 0 & 4 & 0 & $-8$ \\ 
 $S_2$ & $-2$ & $-4$ & 22 & $-2$ & 0 & 4 & $-8$ \\ 
 $S_3$ & $-2$ & $-4$ & 22 & 2 & $-4$ & $-4$ & $-8$ \\ 
 $S_6$ & $-2$ & 10 & 0 & $-2$ & 0 & $-2$ & $-8$ \\ 
 $S_8$ & $-2$ & 10 & 0 & 2 & 2 & 2 & $-8$ \\ 
 $Y_1$ & $-3$ & $-6$ & $-22$ & 0 & $-2$ & 0 & 4 \\ 
 $Y_3$ & $-3$ & $-4$ & 22 & 2 & 2 & 2 & 4 \\ \hline
\end{tabular}

\newpage

\begin{center}
{\Large Table 3} 
\end{center}

\begin{tabular}{|c|l|c|c|c|c|c|c|c|}\hline
SM field & String state & $n_k$ & $\sqrt{5}\hat{Q}'^1$ 
& $\sqrt{55}\hat{Q}'^2$ &
 $\hat{Q}'^5$ & $\sqrt{2}\hat{Q}'^6$ & $\sqrt{2}\hat{Q}'^7$ 
& $\sqrt{3}\hat{Q}'^A$ \\ \hline
$\tilde{q}$ & $Q_L$ & $-1$ & $-12$ & 6  & 0 & 0 & 0 & 0 \\ 
$\tilde{u}$ & $u_L$ & $-1$ & 6 & 42 &  0 & 0 & 0 & 0 \\ 
$\tilde{d}$ & $h_1 d_1-d_2$ & $-2$ & 0 & $-10$  
& ${2(h_1^2 -1) \over 1+h_1^2}$ & ${-4h_1^2 \over 1+h_1^2}$ & 
${4(1-h_1^2) \over 1+h_1^2}$ & 4 \\ 
$\tilde{l}$ & $G_5$ & $-2$ & 2 & 4  & 0 & $-2$ & 0 & $-8$ \\ 
$\tilde{e}$ & $l_5$ & $-2$ & $-4$ & $-8$  & 0 & $-2$ & 0 & $-8$ \\ 
$H$ & $h_1 G_2-G_3$ & $-2$ & 2 & 4  & ${-4h_1^2 \over 1+h_1^2}$
 & ${2(h_1^2 -1) \over 1+h_1^2}$ & 
${2(h_1^2 -3) \over 1+h_1^2}$ & 4 \\ 
$\overline{H}$ & $\bar{G}_1$ & $-1$ & 6 & $-48$  & 0 & 0 & 0 & 0 \\ \hline
\end{tabular}

~~\

\begin{center}
{\Large Table 4} 
\end{center}

\begin{center}
\begin{tabular}{|l|c|c|c|c|c|}\hline
  Rep.& $11a$ & $11b$ & $a+b$ & $\cos^2 \theta$ & $\cos^2 \theta$ \  
(rad. corr.) \\ \hline
  $\tilde{q}$ & 26 & $-37$  & $-1$ & $[0, 0.70]$ & $[0, 0.95]$ \\
  $\tilde{u}$ & 116 & $-193$  & $-7$ & $[0, 0.60]$ & $[0, 0.87]$ \\
  $\tilde{d}$ & $-14$ & 25  & 1 & $[ 0.56,1]$ & [0,1] \\
  $\tilde{l}$ & $-89$ & 144  & 5 & $[ 0.62,1]$ & $[ 0.57,1]$ \\
  $\tilde{e}$ & $-119$ & 196  & 7 & $[ 0.61,1]$ & $[ 0.61,1]$ \\
  $H$ & 76 & $-131$  & $-5$ & --- & --- \\
  $\overline{H}$ & $-109$ & 197  & 8 & --- & --- \\ \hline
\end{tabular}
\end{center}

\end{document}